\documentstyle[prd,aps,psfig]{revtex}
\begin{document}
\draft
\input epsf
\twocolumn[\hsize\textwidth\columnwidth\hsize\csname
@twocolumnfalse\endcsname
%
%
\title{{\hfill \small DAMTP-2001-13, DSF-5-2001, astro-ph/0102144}\\ $~$\\
Early-universe constraints on a time-varying fine structure constant}

\author{P.\ P.\ Avelino${}^{1,2}$, S.\ Esposito${}^{3,4}$, G.\ Mangano${}^{4}$,
C.\ J.\ A.\ P.\ Martins${}^{1,5}$,\\ A.\ Melchiorri${}^{6}$, G.\
Miele${}^{4}$, O.\ Pisanti${}^{4}$, G.\ Rocha${}^{1,6}$, and P.T.P.\
Viana${}^{1,7}$}
\address{$~$\\${}^1$ Centro de Astrof\'{\i}sica, Universidade do Porto,
Rua das Estrelas s/n, 4150-762 Porto, Portugal.}
\address{${}^2$ Dep. de F{\' \i}sica da Faculdade de Ci\^{e}ncias da Univ.
do Porto, Rua do Campo Alegre 687, 4169-007 Porto, Portugal.}
\address{${}^3$ S.I.S.S.A., Via Beirut 2-4, 34014 Trieste, Italy.}
\address{${}^4$ Dipartimento di Scienze Fisiche, Universit\`{a} ``Federico
II'', Napoli, and INFN Sezione di Napoli, Complesso Universitario di Monte
Sant'Angelo, Via Cintia, 80126 Napoli, Italy.}
\address{${}^5$ Department of Applied Mathematics and Theoretical Physics,
Centre for Mathematical Sciences, University of Cambridge\\ Wilberforce
Road, Cambridge CB3 0WA, U.K.}
\address{${}^6$ Department of Physics, Nuclear \& Astrophysics Laboratory,
University of Oxford, Keble Road, Oxford OX1 3RH, U.K.}
\address{${}^7$ Dep. de Matem\'{a}tica Aplicada da Faculdade de
Ci\^{e}ncias da Univ. do Porto, Rua das Taipas 135, 4050 Porto, Portugal.}

\maketitle

\begin{abstract}
{Higher-dimensional theories have the remarkable feature of predicting a
time (and hence redshift) dependence of the `fundamental' four dimensional
constants on cosmological timescales. In this paper we update the bounds on
a possible variation of the fine structure constant $\alpha$ at the time of
BBN ($z\sim10^{10}$) and CMB ($z\sim10^3$). Using the recently-released
high-resolution CMB anisotropy data and the latest estimates of primordial
abundances of $^4He$ and $D$, we do not find evidence for a varying
$\alpha$ at more than one-sigma level at either epoch.}
\end{abstract}

\vskip2pc]

\def\gsim{\;\raise0.3ex\hbox{$>$\kern-0.75em\raise-1.1ex\hbox{$\sim$}}\;}
\def\lsim{\;\raise0.3ex\hbox{$<$\kern-0.75em\raise-1.1ex\hbox{$\sim$}}\;}

\section{Introduction}
\label{secintobs}

In the last few years it has been pointed out that the fundamental energy
scale of gravity does not need to be the Planck scale, but rather it could
be a lower scale maybe not far from the electroweak one
\cite{AHDD,AAHDD,RS}. In this framework, where the hierarchy problem is
definitely solved, the Newton constant turns out to be so small because the
gravitational force spreads into some higher-dimensional space which may be
compact or have an infinite volume. One remarkable feature of
higher-dimensional particle physics theories is that, in this framework,
the coupling constants in the four-dimensional subspace are merely {\it
effective} quantities. Furthermore, from what is known about the dynamics
of these extra dimensions, one expects these effective constants to be time
and/or space varying quantities on cosmological timescales, and this
represents an interesting signature of these models which would be worth to
test. The best example of such a quantity is the fine structure constant
$\alpha$, which is expected to be time-varying in a wide class of theories.

There is quite a large number of experimental constraints on the value of
$\alpha$. These measurements cover a wide range of timescales (see
\cite{alpharev} for a review of this subject), starting from present-day
laboratories ($z \sim 0$), geophysical tests ($z << 1$), and quasars
($z\sim 1 \div 3$), till CMB ($z\sim 10^3$) and BBN ($z\sim10^{10}$)
bounds.

We define $\Delta \alpha/\alpha \equiv \alpha(z)/\alpha-1$, whit $\alpha$
the present value for the fine structure constant. By using atomic clocks
one gets a strong limit on the time variation of the fine structure
constant, $|\Delta \alpha/\alpha| \leq 10^{-14}$ over a period of 140 days
\cite{atomiclocks}. The best geophysical constraint comes from measurements
of isotope ratios in the Oklo natural reactor, which give $|\Delta
\alpha/\alpha| \leq 10^{-7}$ over a period of 1.8 billion years
\cite{geophys}, corresponding to $z \sim 0.1$.

The fine splitting of quasar doublet absorption lines probes higher
redshifts. This is the method of Ref. \cite{alpharev} which finds $\Delta
\alpha/\alpha$=$(-4.6 {\pm} 4.3 $ $({\rm statistical})$ $ {\pm} 1.4 ({\rm
systematic})) 10^{-5}$ for redshifts $z \sim 2 \div 4$, but is subject to
uncertainties associated with laboratory wavelength determinations and
other systematic effects.

On the other hand the analysis of Ref. \cite{Webb} gives a $4\sigma$
evidence for a time variation of $\alpha$, $\Delta
\alpha/\alpha=(-0.72 {\pm} 0.18) 10^{-5}$, for the redshift range $z
\sim 0.5 \div 3.5$. This positive result was obtained using a many-multiplet
method, which is claimed to achieve an order of magnitude greater precision
than the alkali doublet one. Some of the initial ambiguities of the method
have been tackled by the authors with an improved technique, in which a
range of ions is considered, with varying dependence on $\alpha$, which
helps reduce possible problems such as varying isotope ratios, calibration
errors and possible Doppler shifts between different populations of ions
\cite{Quas1,Quas2,Quas3,Quas4}.

A deeper knowledge of the primordial universe at the time of photon
decoupling is emerging from recent results on Cosmic Microwave Background
(CMB) temperature anisotropies
\cite{Boomerang1,Boomerang2,Maxima1,Maxima2,Dasi}, and on this ground, a
new bound on $|\Delta \alpha/\alpha|$ at $z\sim10^3$ has been obtained
\cite{me00,oth} by using the first release of data by BOOMERanG and MAXIMA
\cite{Boomerang1,Maxima1}. As observed by many authors, the analysis of
cosmological implications of this first bunch of data seemed to prefer
values for the baryon fraction sensibly larger than the Big Bang
Nucleosynthesis (BBN) requirement \cite{Emmp2,cris1,cris2}. This
preliminary analysis thus suggested that likely {\it new physics} was
active in the early universe, and in this direction many different
mechanisms have been proposed in the literature
\cite{Emmp2,cris1,cris2,cris3,cris4,cris5,other1,other2,other3,other4,other5,other6}.

In this paper we update the constraints on $|\Delta
\alpha/\alpha|$ from CMB by using the latest available observational data
\cite{Boomerang2,Dasi} which now single out a value for the baryon
fraction in perfect agreement with the BBN one \cite{HMMMP}.

At higher redshift, $z\sim10^{10}$, BBN can provide strong bounds on a
possible deviation of the value of fine structure constant from the
present-day one. In this paper we obtain new constraints on $|\Delta
\alpha/\alpha|$ at the time of BBN, based on the new data on primordial
chemical abundances and on a new and more precise BBN code recently developed
\cite{Emmp1,Emmp2}. This represents an update of the analysis of Ref.
\cite{Rubinstein}.

\section{BBN data analysis}
\label{bbn}

The high level of predictivity of {\it Standard} nucleosynthesis (SBBN),
which yields abundances for $D$, $^3He$, $^4He$ and $^7Li$ as function of
the baryon fraction $\Omega_b h^2$ only\footnote{We assume here the
standard scenario with only the three active neutrinos, and photons,
contributing to the relativistic energy density.}, makes the comparison of
the theoretical BBN predictions with experimental data a crucial test for
Hot Big Bang models.

The measurements of the deuterium Ly-$\alpha$ features in several Quasar
Absorption Systems at high red-shift ($z > 2$) give a relative deuterium
abundance $D/H = (3.0 {\pm} 0.4){\times}10^{-5}$ \cite{Tytler}. Deuterium
has a relevant role in BBN since it mainly fixes the baryon
fraction. For the $^4He$ mass fraction, $Y_P$, the key results come from
the study of HII regions in Blue Compact Galaxies. This has been performed
in two different analyses, which give rather different values for
$Y_P$. The value found in ref.\cite{Olive} is $Y_P = 0.234 {\pm} 0.002$
and it is quite smaller than $Y_P = 0.244 {\pm} 0.002$ obtained in
Ref.\cite{IzotovThuan}. However, even though both analyses use large
samples of objects, the analysis of Ref.\cite{IzotovThuan} is based on a
more homogeneous set of measurements. Moreover, some of the most metal-poor
objects used in \cite{Olive} seem to suffer from stellar absorption. For
these reasons we use the value $Y_P = 0.244 {\pm} 0.002$.

Inferring the $^7Li$ primordial abundance is a rather difficult task, since
it is strongly affected by stellar processes. The value reported in Ref.
\cite{Ryan}, $^7Li/H= 1.23^{+0.68}_{-0.32}{\times}10^{-10}$, is based on the
measurement of $^7Li$ in the halos of old stars. It should represent well
its primordial abundance since in Ref. \cite{Ryan} it is also taken into
account production and depletion mechanisms due to cosmic rays and stellar
dynamics, respectively. This result, which is compatible with other similar
analysis \cite{Li7a,Li7b,Li7c} is unfortunately smaller by a factor 2 to 3
than what is typically predicted by BBN. Moreover, stellar models where
strong $^7Li$ depletion mechanisms are present have been extensively
discussed in the literature and are also supported by the observation of
old stars where no $^7Li$ at all is present in the halo. For this reason,
at the moment the $^7Li$ primordial abundance cannot be safely used in a
BBN analysis in order to impose bounds on the baryon fraction or other
parameters.

The predictions of SBBN \cite{Emmp2,FLSV,LSV}, for three standard light
neutrinos, $N_\nu=3$, once compared with the above experimental
observations through a likelihood analysis, yield a baryon fraction which
is in the range $\Omega_b h^2 = 0.019^{+0.004}_{-0.002}$ (at $95\%$ C.L.),
in fair agreement with the value $\Omega_b h^2 = 0.020 {\pm} 0.002$ (at $95\%$
C.L.) of Ref. \cite{Burles}\footnote{Note that our wider range for the
baryon fraction is basically due to the larger uncertainty on the deuterium
abundance we use in our analysis (see Ref.\cite{Tytler}) with respect to
the experimental data used in \cite{Burles}.}. SBBN for the central value
$\Omega_b h^2 = 0.019$ gives $D/H = 3.26{\times}10^{-5}$, $Y_P= 0.2467$ and
$^7Li/H= 3.31{\times}10^{-10}$, which corresponds to $\chi^2 = 2.1$.

A variation of the value of the fine structure function $\alpha$ does not
have negligible effects on SBBN abundance predictions. This issue has been
already investigated in the literature \cite{Rubinstein} in order to fix
bounds on $\Delta \alpha/\alpha$ at $z\sim10^{10}$. The effect on BBN of a
varying $\alpha$ is essentially twofold, affecting both the neutron-proton
mass difference and the Coulomb barrier in nuclear reactions. The mass
difference between neutron and proton, $\Delta m$, since it fixes the
neutron to proton ratio at decoupling, provides the initial condition at
the onset of BBN. The $\alpha$ dependence of $\Delta m$ can be derived
phenomenologically, as done in Ref. \cite{Gasser}, $\Delta m \simeq 2.05 -
0.76 ( 1 + {\Delta
\alpha / \alpha })$ MeV, whereas the dependence on $\alpha$ of the most
important nuclear reactions involved in BBN has been carefully evaluated
and reported in Table 1 of Ref. \cite{Rubinstein}. Both effects have been
implemented in a high accuracy BBN code \cite{Emmp2,Emmp1}, to produce the
light element abundances as functions of $\Omega_b h^2$ and $\Delta
\alpha/\alpha$.

It is important to mention that, in general, if we consider models where
the electromagnetic coupling is a time dependent parameter, it is
reasonable to expect that $all$ fundamental parameters, the Yukawa
couplings, the strong coupling constant and Weinberg angle, and the vacuum
expectation value (vev) for the Higgs field, $v$, may be functions of time
as well. In particular, a different value for the Fermi constant, $G_F\sim
v^{-2}$\footnote{Notice that $G_F$ does not depend on the electroweak
couplings.}, would change all rates of weak processes and result in
different BBN predictions. Nevertheless we have chosen to keep constant the
values for all these parameters. This represents the simplest scenario
which may account for Quasar measurements and provides the most restrictive
bounds on $\Delta \alpha /\alpha$. On the other hand, since, in the general
case, the time dependence of fundamental parameters is model dependent, a
completely general analysis clearly loses any predictivity.

By using the BBN results on D and $^4He$ abundances we have performed a
likelihood analysis in the plane $\Omega_b h^2 - \Delta \alpha/\alpha$. The
68\% and 95$\%$ C.L. contours are reported in Figure 1. By marginalizing
with respect to $\Omega_b h^2$ and $\Delta \alpha/\alpha$ one gets the
allowed intervals at 95$\%$ C.L., $\Omega_b h^2 =0.020^{+0.005}_{-0.003}$
and $\Delta \alpha/\alpha = (-7{\pm}9){\times}10^{-3}$, respectively. This result,
which is compatible with the bound found in \cite{Rubinstein}, shows that
BBN does not clearly favour a value of $\Delta
\alpha/\alpha
\neq 0$ at more than $1-\sigma$ level. For the maximum of the likelihood we
find $D/H = 2.98{\times}10^{-5}$, $Y_P= 0.2440$ and $^7Li/H= 3.86{\times}10^{-10}$, with
a $\chi^2 = 1.3{\times}10^{-3}$. The agreement with experimental observation is
much improved, due to the additional free parameter $\Delta \alpha
/\alpha$.

\section{CMB data analysis}
\label{secdatres}

We have performed a similar likelihood analysis for the recently released
BOOMERanG \cite{Boomerang2} and DASI \cite{Dasi} data, as well as the COBE
data, with the additional free parameter $\Delta \alpha/ \alpha$. Our
analysis method follows the procedure described in \cite{deb01} taking into
account the effects of the beam and calibration uncertainties for the
Boomerang data. For the DASI data we consider the public available
correlation matrices and window functions.

\begin{figure}
\vbox{
\hbox{
\psfig{file=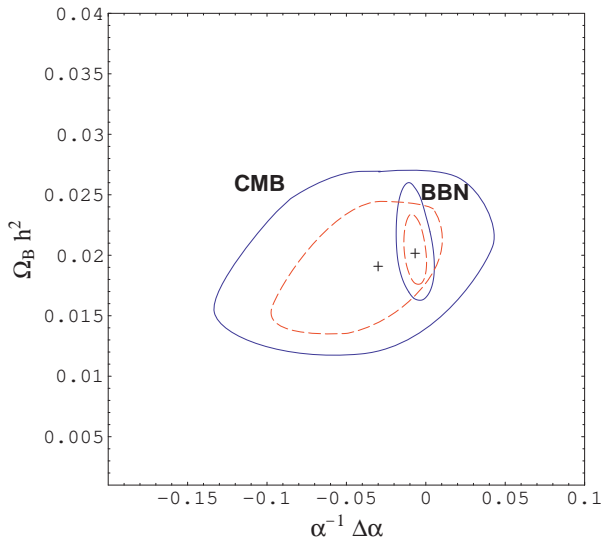,width=8cm}}}
\caption{The dashed and solid lines represent the $68 \%$ and $95 \%$ C.L.
contours, respectively, for the CMB and BBN likelihood analysis. Crosses
correspond to the maxima of the likelihood functions.}
\end{figure}

The calibration uncertainty is taken into account by adding a gaussian term
$\chi^2_{cal}=(1-A_{cal})^2/\sigma_{cal}^2$ to the computed $\chi^2$  for
each theoretical spectrum. $A_{cal}$  is a calibration parameter and
$\sigma_cal=0.23$, $0.08$ for Boomerang and DASI respectively.

The calculation of the angular power spectrum $C_l$ follows
\cite{me00,steen,Kap,recfast,recfasta} and was obtained using a modified
CMBFAST algorithm which allows a varying $\alpha$ parameter. The space of
model parameters spans $\Omega_m=(0.1 - 1.0)$, $\Omega_{b}h^{2}=(0.009 \div
0.036)$, $h=(0.4 \div 0.9)$, $ \Delta \alpha / \alpha=(-0.2 \div 0.1)$,
$n_s=(0.7
\div 1.3)$. The basic grid of models was obtained considering parameter step
sizes of 0.1 for $\Omega_m$; 0.003 for $\Omega_{b}h^{2}$; 0.05 for $h$;
0.01 for $\Delta \alpha / \alpha$ and finally 0.05 for the tilt $n_s$. When
necessary the grid resolution is increased by using interpolation
procedures. All our models have $\Omega_{total}=1$. We assume the presence
of a classical cosmological constant when necessary to achieve such result.
We also assume an age of the universe prior $t_0>10 Gyr$.

Performing the marginalization over one of the two parameters $\Delta
\alpha / \alpha$ and $\Omega_b h^2$ gives respectively $\Omega_b h^2
= 0.020^{+0.004}_{-0.004}$ and $\Delta
\alpha / \alpha = -0.05^{+0.07}_{-0.04}$ at $68 \%$ C.L.. The CMB results
on $\alpha$ can be further constrained by the inclusion of external
priors on cosmological parameters. Assuming $h=0.72 {\pm}
0.08$ \cite{freeman}, for example, yields $\Delta\alpha / \alpha =
-0.02^{+0.03}_{-0.04}$, while assuming $\Omega_b h^2 =
0.019^{+0.004}_{-0.002}$ gives $\Delta\alpha / \alpha = -0.03 {\pm} 0.05$. To
compare CMB results with BBN constraints on $\Delta \alpha/ \alpha$ we use
this latter BBN prior result. This is perfectly justified in view of the
excellent agreement between the two determination of $\Omega_b h^2$ from
CMB and BBN. The $68 \%$ and $95 \%$ C.L. regions in the plane $\Omega_b
h^2$
-- $\Delta
\alpha / \alpha$ are shown in Fig. 1.

\section{Conclusions}
\label{secdiscon}

The analysis of the $\alpha$-dependence of two relevant cosmological
observables like the anisotropy of CMB and the light element primordial
abundances does not support evidence for variations of the fine structure
constant at more than the one-sigma level at either epoch. This is the
first time a joint analysis for the two epochs has been done, and as such
it is quite a robust result.

A few comments are nevertheless in order. The most noticeable aspect of our
results is the apparent disagreement with earlier work of some of the
present authors \cite{me00}. However, the discrepancy is trivially
explained by the use of different CMB datasets. Indeed, the earlier release
of BOOMERanG and MAXIMA \cite{Boomerang1,Maxima1} data which was used by
\cite{me00} singled out a quite large value for the baryonic fraction with
respect to the BBN prediction. Thus, in that scenario a smaller value of
$\alpha$ with respect to the present-day one, was a possible way to lower
the value of $\Omega_b h^2$, making it compatible with BBN.

In the new release of data from BOOMERanG and DASI, this baryon discrepancy
has been eliminated, at the price of a sligtly lower spectral index
($n_s\sim0.9$ as opposed to the previously preferred $n_s\sim1.00$). In
this context, and given the intrinsic degeneracies in the problem, a
slightly negative $\alpha$ while still able to marginally improve the fits
is no longer a significant advantage. As was emphasized in \cite{me00},
more data and an independent knowledge of other cosmological parameters
will be needed in order to obtain a more precise 'measurement' of $\alpha$
from the CMB. We point out, however, that a recent re-analysis of the old
Maxima-I dataset with an increased pixel resolution produced results still
in better agreement with an high baryon fraction \cite{Maxima2}.

Hence, from the observational viewpoint, the only current strong evidence
for a varying $\alpha$ seems to be the four-sigma detection using quasar
data at redshifts $z\sim 1 - 3$. It should be said that even imposing
fairly strong constraints at redshift $z\sim10^{10}$ and $z\sim10^3$, our
results cannot strictly be extrapolated for the whole cosmological period
in between these epochs. Indeed, two-metric models exist where $\alpha$ and
other constants suffer `temporary' variations for fairly limited time
periods, the case in point being the epoch of equal matter and radiation
densities.

What our results, together with the quasar data, do strongly rule out is
any cosmological model where $\alpha$ behaves as a simple and smooth
power law function of say the scale factor or cosmic time.
If there were indeed any variations of $\alpha$ in the past, then they
are likely to have been fairly `sharp', most likely as a side-effect
of phase transitions or other dramatic events in the history of the universe.

\acknowledgements

We thank J. Barrow, R. Caldwell, B. Carter, G. Esposito-Farese,
D. Langlois and S. Sarkar for useful comments.
C.\ M.\ is funded by FCT (Portugal) under ``Programa PRAXIS XXI'' (grant no.\
FMRH/BPD/1600/2000). We thank Centro de Astrof{\'{\i}}sica da Universidade
do Porto (CAUP) for the facilities provided. G. R.
also thanks the Dept. of Physics of the University of Oxford for support
and hospitality during the progression of this work.
G.R. and C.M. are funded by FCT (Portugal).


\end{document}